\documentclass[default]{sn-jnl}

\jyear{2022}%

\theoremstyle{thmstyleone}%
\theoremstyle{thmstyletwo}%

\theoremstyle{thmstylethree}%

\usepackage{hyperref}

\usepackage{tabularx}
\usepackage{etoolbox}
\usepackage{booktabs}
\usepackage{graphicx}
\usepackage{tabu,multirow}
\usepackage{colortbl,dcolumn}
\makeatletter
\patchcmd{\@makecaption}
\renewcommand\thetable{\thesection-\arabic{table}}
{\scshape}

\usepackage{listings}
\usepackage{xcolor}
\definecolor{codegreen}{rgb}{0,0.6,0}
\definecolor{codegray}{rgb}{0.5,0.5,0.5}
\definecolor{codepurple}{rgb}{0.58,0,0.82}
\definecolor{backcolour}{rgb}{0.95,0.95,0.92}
\lstdefinestyle{mystyle}{
    backgroundcolor=\color{backcolour},   
    commentstyle=\color{codegreen},
    keywordstyle=\color{magenta},
    numberstyle=\tiny\color{codegray},
    stringstyle=\color{codepurple},
    basicstyle=\ttfamily\footnotesize,
    breakatwhitespace=false,         
    breaklines=true,                 
    captionpos=b,                    
    keepspaces=true,                 
    numbers=left,                    
    numbersep=5pt,                  
    showspaces=false,                
    showstringspaces=false,
    showtabs=false,                  
    tabsize=2
}
\begin{document}
	
	
	\title[ ]{NodeCoder: a graph-based machine learning platform to predict active sites of modeled protein structures.} 
	
	
	\author[1,2]{\fnm{Nasim} \sur{Abdollahi}}\email{nasim.abdollahi@utoronto.ca}
	\author[2]{\fnm{Seyed Ali} \sur{Madani Tonekaboni}}\email{ali.madani@cyclicarx.com}
	
	\author[2]{\fnm{Jay} \sur{Huang}}\email{jingjing.huang@cyclicarx.com}
	
	\author*[1,3,4,5]{\fnm{Bo} \sur{Wang}}\email{bowang@vectorinstitute.ai}
	
	\author*[2]{\fnm{Stephen} \sur{MacKinnon}}\email{stephen.mackinnon@cyclicarx.com }
	
	\affil*[1]{\orgdiv{Department of Laboratory Medicine and Pathobiology}, \orgname{University of Toronto}, \orgaddress{\city{Toronto}, \state{ON}, \country{Canada}}}
    
	\affil[2]{\orgname{Cyclica Inc.}, \orgaddress{\city{Toronto}, \state{ON}, \country{Canada}}}
	
	\affil[3]{\orgname{Vector Institute for Artificial Intelligence}, \orgaddress{\city{Toronto}, \state{ON}, \country{Canada}}}
	
	\affil[4]{\orgname{Peter Munk Cardiac Center, University Health Network}, \orgaddress{\city{Toronto}, \state{ON}, \country{Canada}}}
	
	\affil[5]{\orgname{Department of Computer Science}, \orgaddress{\city{Toronto}, \state{ON}, \country{Canada}}}
	
	
	\abstract{While accurate protein structure predictions are now available for nearly every observed protein sequence, predicted structures lack much of the functional context offered by experimental structure determination. We address this gap with NodeCoder, a task-independent platform that maps residue-based datasets onto 3D protein structures, embeds the resulting structural feature into a contact network, and models residue classification tasks with a Graph Convolutional Network (GCN).  We demonstrate the versatility of this strategy by modeling six separate tasks, with some labels derived from other experimental structure studies (ligand, peptide, ion, and nucleic acid binding sites) and other labels derived from annotation databases (post-translational modification and transmembrane regions).  Moreover, A NodeCoder model trained to identify ligand binding site residues was able to outperform P2Rank, a widely-used software developed specifically for ligand binding site detection. NodeCoder is available as an open-source python package at \href{https://pypi.org/project/NodeCoder/}{https://pypi.org/project/NodeCoder/}.}
	
	
	
	\maketitle
	
	\section{Introduction}\label{sec1}
	%
    
    Protein 3D structure prediction has undergone a sudden leap in predictive performance with the emergence of deep-learning models \cite{pereira2021high, jumper2021highly, baek2021accurate}.  In turn, reliable template-free structure prediction can now be applied at scale to model full proteomes. A partnership between DeepMind and the EMBL-EBI published over 200 million predicted protein structures predicted by AlphaFold2 and provided open access to accelerate scientific research \cite{DeepMind_3DHuman, AlphaFold2}.  This open science initiative was followed by Meta, who later released predicted structures for 617 million protein structures from metagenomics sources, built using their structure prediction tool built off protein language-models \cite{Lin2022.07.20.500902}.  For reference, protein structures determined by experimental means can take months to years to solve. As of November 11th, 2022 the Protein DataBank (PDB) lists 197,848 experimentally determined structures deposited over its 50 years history \cite{PDB}.  
    
    These widely accessible and high-accurate protein structures have the potential to democratize drug discovery opportunities.  Countless protein systems are already available on demand, while others can be rapidly generated.  For example, the emerging genomic sequences of new pathogens are typically available within days or weeks of new outbreaks and are subsequently ready for 3D protein structure predictions. Following the publication of the novel coronavirus genome in January 2020, reliable 3D protein structure models for viral proteins were automatically generated and made available within days using homology models and \textit{de novo} prediction strategies, including SwissModel\cite{SWISSMODEL_SARSCoV2, SIB_coronavirus}, ZhangLab \cite{ZhangLab_SARSCoV2, zhang2020protein}, AlphaFold \cite{DeepMind_COVID19} and Rosetta \cite{Rosetta_coronavirus}.  
    
    Despite the large recent gains in structure prediction accuracy, using predicted structures effectively for pharmaceutical applications remains a challenge. First, predicted protein structures may have highly accurate topology but lack the atomistic detail of X-Ray crystal structures near critical binding surfaces \cite{Terwilliger2022.11.21.517405}.  Most computational and AI technologies developed for drug discovery are designed to operate on single, high-resolution X-ray crystal structures with positional accuracy down to the atomistic level.  Since predicted protein structures lack the positional accuracy and multiple conformations provided by experimental means, there is an emerging need for non-atomistic solutions that can drive new drug discovery insights.  Secondly, modeled protein structures lack the added structural context offered through experimental means, such as ligand binding sites, post-translational modifications, macromolecular binding surfaces, metal binding sites, solvent binding sites, etc.  
    
    We address these challenges with NodeCoder, a graph-based deep learning model for arbitrary residue classification tasks.  This framework maps any residue-based datasets onto their predicted protein structures, generates salient biophysical features, and embeds each residue into a graph representation of its contact network to train a graph convolutional network (GCN) model, Fig. \ref{Schematic Representation}. GCNs are capable of capturing complex relationships, making them especially suitable to a broad range of problems within computational biology, structural biology and computational drug discovery \cite{morselli2021network, zeng2020repurpose, lin2022generalizeddta, stark2022equibind, you2021deepgraphgo}. Within a GCN, each node is represented as its own feature vector, while convolutional layers facilitate information exchange between neighboring nodes. In turn, additional convolutional layers connect nodes across larger graph distances, allowing the model to learn higher-order information. In the context of a 3D protein structure, residue contact networks are naturally adaptable to graph-based convolutional neural networks, which have been shown to outperform CNN-based architectures on protein (graph) classification tasks \cite{zamora2019structural}. 
    
    NodeCoder is a GCN-based framework designed to model node (residue) classification tasks and used to predict key structural and functional sites on AlphaFold2 protein structures.  Previously, different categories of structural or functional sites were each considered their own distinct problems, each in need of dedicated expert-designed solutions \cite{ahmad2005pssm, yan2006predicting, wang2006bindn, petsalaki2009accurate, krivak2018p2rank, sodhi2004predicting, li2008prediction, chen2021structure}.  In this study, we apply a single generalizable graph-based learning framework built to six diverse predictive tasks whose residue-based labels originate from experimental protein structures (ligand, peptide, nucleic acid, and metal binding sites) and non-structure based sources (modified residues and transmembrane domains).  Unlike predictive engines based on primary sequence alone, the proposed graph-based methodology benefits from the large structural coverage of proteins in the open-source dataset and introduces spatial and local environment context to all predictions.  Lastly, we demonstrate that the universal modeling platform can achieve predictive performance comparable to leading task-specific predictive solutions.

    \section{Results}\label{Results}

\begin{figure}
		\centering
		\includegraphics[width=1.0\linewidth]{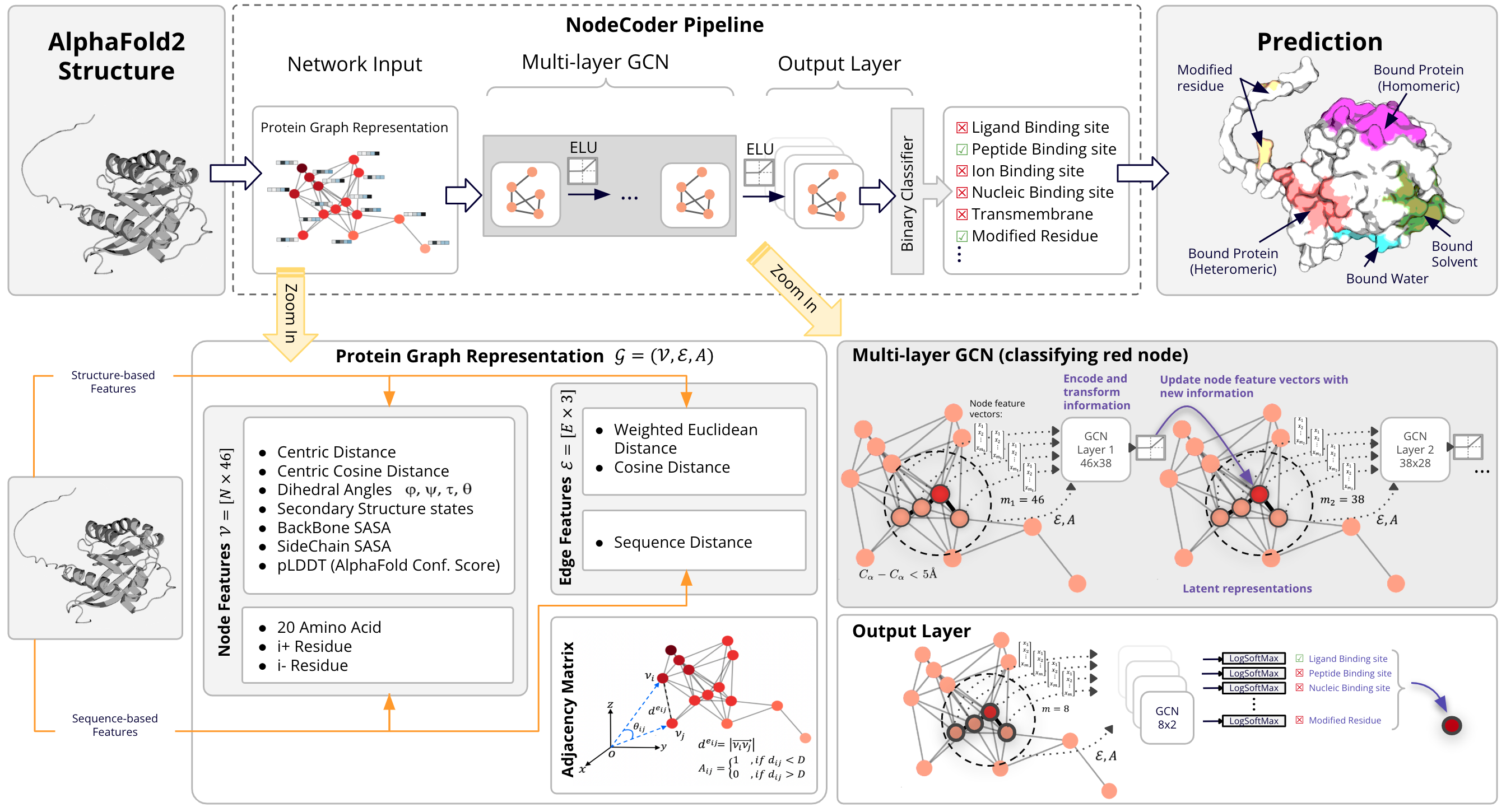}
		\caption{Schematic representation of NodeCoder platform based on a graph convolutional network model that embeds representation of an AlphaFold2 predicted protein structure and performs residue classification tasks. The protein residue graph is constructed from its 3D structure, where nodes correspond to individual amino acid residues and edges correspond to inter-residue contact within a pre-set distance denoted. Node features, the known properties of the residues, are derived from the protein's 3D structure model. NodeCoder annotates the residues through the exploitation of structure-based and sequence-based features of the target residue and its neighbors in the protein graph. NodeCoder architecture consists of upstream shared multi-layer GCN, stacked GCNs, and as many independent GCN modules as the number of output tasks.}
		\label{Schematic Representation}
	\end{figure}

    The goal of the proposed unified NodeCoder framework is to accept an input protein structure exclusively made up of residue coordinates and annotate specific regions of interest based on the structural attributes of the residue itself and its local neighbors. NodeCoder maps residue-based datasets onto predicted protein structures in order to augment learning with 3D spatial context and subsequently apply the trained models for inference on new predicted structures (Fig. \ref{Schematic Representation}).  To reduce data source bias on inference, AlphaFold2 (AF2) structures were used for training on all tasks, including those whose labels originate from experimental structures.  Moreover, the study was limited to human proteins and one 3D representation per protein to avoid performance over-estimations attributed to orthologs or other forms of test/train split data leakage. Proteins lacking any binding annotations are not used for training to prevent excessive false negative labels from biasing the dataset, leaving $3823$ total proteins. The $3823$ proteins are randomly split into five cross-validation folds. Additionally, five separate protein graphs are built with five different distance thresholds of $5\textrm{\AA}$, $8\textrm{\AA}$, $10\textrm{\AA}$, $12\textrm{\AA}$ and $14\textrm{\AA}$ between residue nodes for each cross-validation fold. Two residues (nodes) have an edge in the protein graph if the distance between their corresponding $C_\alpha$ atoms is within the specified threshold. The distance threshold defines the residues connections and ensures different structural levels are captured.  Increasing the distance threshold yields a protein graph with a higher node degree and overall graph connectivity.  The increased connectivity allows each convolutional layer to exchange information through larger spatial distances. The five distance thresholds of $5\textrm{\AA}$, $8\textrm{\AA}$, $10\textrm{\AA}$, $12\textrm{\AA}$ and $14\textrm{\AA}$ resulted in similar node-degree distributions across all train and validation folds with the average node-degree of $2$, $8$, $14$, $23$ and $33$ edges respectively.  
	
	\subsection{Modeling Biophysical Recognition Tasks}\label{Observation1}

    To demonstrate NodeCoder's task-independent architecture, we prepare six residue classification tasks by mapping residue annotations from multiple BioLip and UniProt entries onto their corresponding protein dataset (Table \ref{annotation list}). These tasks include post-translational modifications and transmembrane spanning residues from UniProt \cite{uniprot, uniprot2021uniprot}, as well as ligands, nucleic acid, peptide, and metal ion binding sites mapped from BioLip \cite{yang2012biolip}. The total node count and percentage of positive labels in each train and validation fold are summarized in Table \ref{Summary_of_Folds}. 
    
\begin{figure*}[!ht]%
		\centering
		\includegraphics[width=1\textwidth]{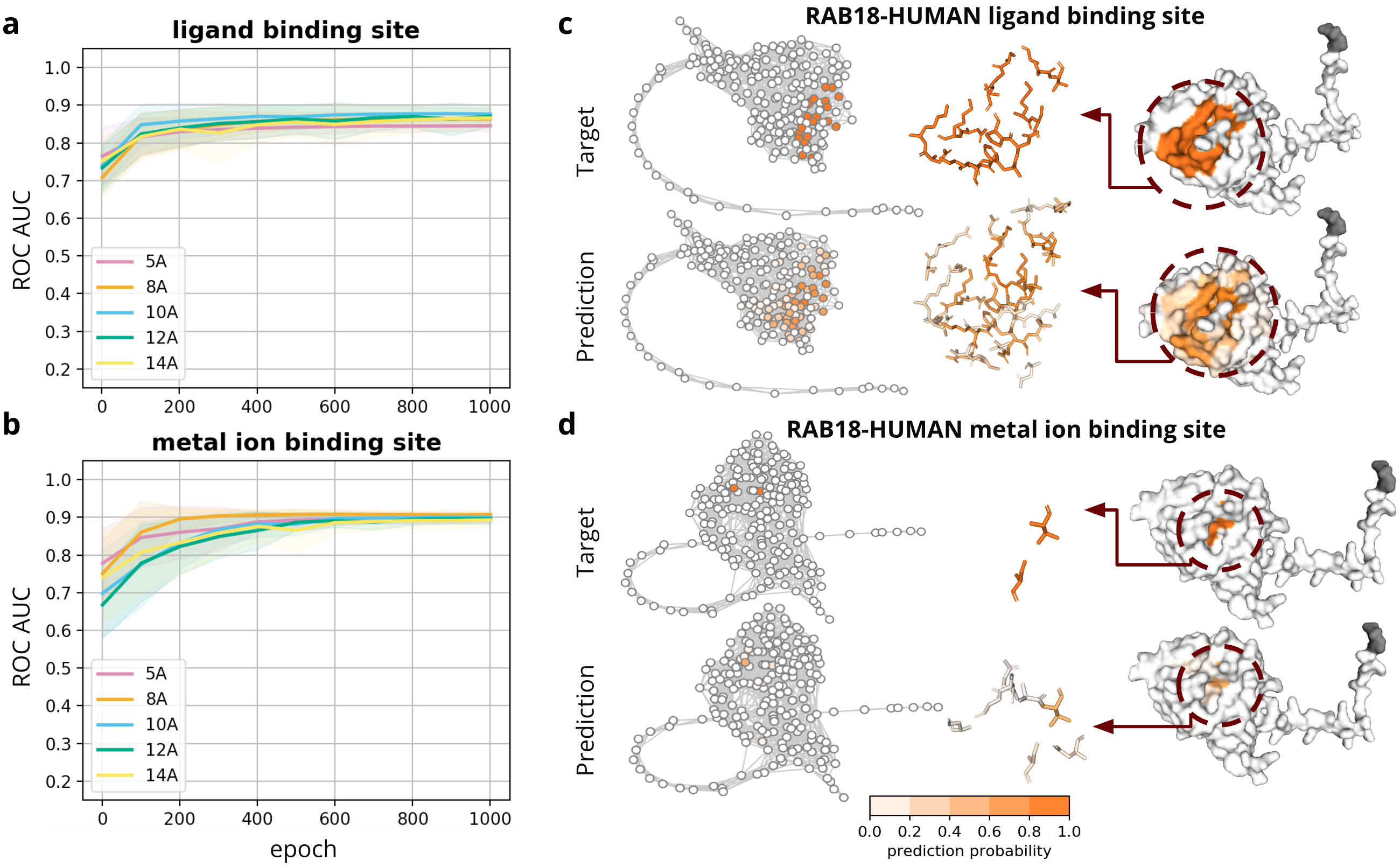}
		\caption{Left: ROC AUC for two classification tasks, (a) ligand binding site and (b) metal ion binding site, each generated by five separate models at varying distance thresholds for the protein contact network ($5\textrm{\AA}$, $8\textrm{\AA}$, $10\textrm{\AA}$, $12\textrm{\AA}$, and $14\textrm{\AA}$). The solid line is the average on five cross-validation folds and the shaded area illustrates the average $\pm$ standard deviation. 2D and 3D representations of true binding sites (Target) and predicted binding sites (Prediction) of RAB18-HUMAN: (c) ligand binding site and (d) metal ion binding site. In the 2D representation, each circle is a single amino acid residue in the protein graph centered at its $C_\alpha$ coordinate. The stick representation illustrates the main and side chain of the amino acid residues of the binding site and its surrounding. The color intensity in ``Prediction" indicates the prediction probability of the residue being classified with the downstream task. The 3D surface protein visualization and the stick representation of the amino acid residues are created using PyMOL.}\label{ROCAUC_Bindingtasks_RAB18_Human}
	\end{figure*}

    To quantify the network performance, the area under receiver operating characteristic curve (ROC AUC), with the ``True Positive Rate" on the vertical axis and the ``False Positive Rate" on the horizontal axis, is calculated once each epoch is completed. ROC AUC is then averaged on five validation folds for every prediction experiment. Fig. \ref{ROCAUC_Bindingtasks_RAB18_Human} - (a) and (b) show the averaged ROC AUC measure of five validation fold for NodeCoder models varying in distance threshold for two prediction tasks of ligand binding site and metal ion binding site. In addition to ROC AUC, area under the precision-recall curve (PR AUC) is also calculated for all five validation folds after each training epoch. The ROC AUC and PR AUC scores of the validation folds are averaged and the best scores obtained for NodeCoder models trained with five different distance thresholds are summarized in Table \ref{Summary_of_performance}.

    We found that the optimal contact distance threshold varies from task to task.  Specifically, predictive models for large macromolecular interaction surfaces prefer large contact distance thresholds. Transmembrane residues, which represent lipid bi-layer interactions, favored the largest measured distance threshold at 14$\textrm{\AA}$, followed by nucleic acid and peptide binding sites at 12$\textrm{\AA}$, ligand binding at 10$\textrm{\AA}$, metal ion binding at 8$\textrm{\AA}$, and lastly, the modified residue predictions favored the 5$\textrm{\AA}$ cutoff.  Moreover, Figure \ref{ROCAUC_Bindingtasks_RAB18_Human} reports faster learning rates at the preferred distance threshold for the ligand and metal ion binding models in panels (a) and (b) respectively.  Panels (c) and (d) illustrate 2D and 3D representations of the actual ligand and ion binding sites, ``Target'', and predicted binding sites, ``Prediction'', for RAB18-HUMAN.  The color intensity in ``Prediction'' indicates the prediction probability of the residue on its given task. Residues near the center of experimentally-observed binding sites have stronger predictive signals, while those at the edge of the binding surface are predicted with lower confidence (Figure \ref{ROCAUC_Bindingtasks_RAB18_Human} (c)).  
    
    Task difficulty was also highly variable.  Predicting transmembrane regions from a 3D structure is nearly trivial given the consistent structural topology and residue properties of membrane-spanning regions.  Transmembrane residue prediction reached a ROC AUC of 0.993 and this relatively easy task served as a positive control in the design and development of NodeCoder.  Peptide binding site predictions were the most difficult for NodeCoder, reaching a ROC AUC performance of $0.813$.  Peptide-binding surface predictions are considerably more difficult as the structural determinants that differentiate peptide interactions from other ubiquitous forms of protein self-interaction, homomeric or heteromeric interactions, may be very subtle.  In contrast, nucleic acid binding surfaces have well-defined structural determinants, such as cationic residues to offset charged phosphates \cite{Nadassy1999}, and were also easier for NodeCoder to recognize (ROC AUC = 0.909).  
	
	\begin{table*}[htb]
		\centering
		\caption{Average of the best NodeCoder ROC AUC and PR AUC scores across validation fold trained with different distance thresholds.}
		\label{Summary_of_performance}
		\scalebox{0.48}{\begin{tabular} {@{}ccccccccccccl@{}}
				\midrule 
				\parbox{4.2cm}{\textbf{Distance thresholds:}} & \multicolumn{2}{c}{\parbox{2cm}{\centering \textbf{$5\textrm{\AA}$}}} & \multicolumn{2}{c}{\parbox{2cm}{\centering \textbf{$8\textrm{\AA}$}}} & \multicolumn{2}{c}{\parbox{2cm}{\centering \textbf{$10\textrm{\AA}$}}} & \multicolumn{2}{c}{\parbox{2cm}{\centering \textbf{$12\textrm{\AA}$}}} & \multicolumn{2}{c}{\parbox{2cm}{\centering \textbf{$14\textrm{\AA}$}}}\\ \midrule
				
				\parbox{1cm}{\centering \textbf{}} & \parbox{1.9cm}{\centering \textbf{ROC AUC}} & \parbox{1.6cm}{\textbf{PR AUC}} & \parbox{1.9cm}{\centering \textbf{ROC AUC}} & \parbox{1.6cm}{\textbf{PR AUC}} & \parbox{1.9cm}{\centering \textbf{ROC AUC}} & \parbox{1.6cm}{\textbf{PR AUC}} & \parbox{1.9cm}{\centering \textbf{ROC AUC}} & \parbox{1.6cm}{\textbf{PR AUC}} & \parbox{1.9cm}{\centering \textbf{ROC AUC}} & \parbox{1.6cm}{\textbf{PR AUC}} \\ \midrule
				
				\parbox{4.2cm}{\textbf{ligand binding site}} 
				& \parbox{1.5cm}{\centering{\cellcolor{orange!10}0.849}} & \parbox{1.5cm}{\centering{\cellcolor{orange!10}0.257}}
				& \parbox{1.5cm}{\centering{\cellcolor{orange!20}0.873}} & \parbox{1.5cm}{\centering{\cellcolor{orange!20}0.288}}
				& \parbox{1.5cm}{\centering{\cellcolor{orange!50}\textbf{0.884}}} & \parbox{1.5cm}{\centering{\cellcolor{orange!50}\textbf{0.295}}}
				& \parbox{1.5cm}{\cellcolor{orange!40}\centering{0.882}} & \parbox{1.5cm}{\cellcolor{orange!40}\centering{0.275}}
				& \parbox{1.5cm}{\cellcolor{orange!30}\centering{0.879}} & \parbox{1.5cm}{\cellcolor{orange!30}\centering{0.256}}\\ 
				
				\parbox{4.2cm}{\textbf{peptide binding site}}
				& \parbox{1.5cm}{\cellcolor{orange!10}\centering{0.769}} & \parbox{1.5cm}{\cellcolor{orange!10}\centering{0.039}}
				& \parbox{1.5cm}{\cellcolor{orange!20}\centering{0.789}} & \parbox{1.5cm}{\cellcolor{orange!20}\centering{0.046}}
				& \parbox{1.5cm}{\cellcolor{orange!30}\centering{0.793}} & \parbox{1.5cm}{\cellcolor{orange!30}\centering{0.047}}
				& \parbox{1.5cm}{\cellcolor{orange!50}\centering{\textbf{0.813}}} & \parbox{1.5cm}{\cellcolor{orange!50}\centering{\textbf{0.054}}}
				& \parbox{1.5cm}{\cellcolor{orange!40}\centering{0.810}} & \parbox{1.5cm}{\cellcolor{orange!40}\centering{0.050}}\\ 
				
				\parbox{4.2cm}{\textbf{nucleic acid binding site}}
				& \parbox{1.5cm}{\cellcolor{orange!10}\centering{0.869}} & \parbox{1.5cm}{\cellcolor{orange!10}\centering{0.086}}
				& \parbox{1.5cm}{\cellcolor{orange!30}\centering{0.89}} & \parbox{1.5cm}{\cellcolor{orange!30}\centering{0.128}}
				& \parbox{1.5cm}{\cellcolor{orange!20}\centering{0.886}} & \parbox{1.5cm}{\cellcolor{orange!20}\centering{0.119}}
				& \parbox{1.5cm}{\cellcolor{orange!50}\centering{\textbf{0.909}}} & \parbox{1.5cm}{\cellcolor{orange!50}\centering{\textbf{0.157}}}
				& \parbox{1.5cm}{\cellcolor{orange!40}\centering{0.895}} & \parbox{1.5cm}{\cellcolor{orange!40}\centering{0.128}}\\ 
				
				\parbox{4.2cm}{\textbf{metal ion binding site}}
				& \parbox{1.5cm}{\cellcolor{orange!10}\centering{0.900}} & \parbox{1.5cm}{\cellcolor{orange!10}\centering{0.256}}
				& \parbox{1.5cm}{\cellcolor{orange!50}\centering{\textbf{0.915}}} & \parbox{1.5cm}{\cellcolor{orange!50}\centering{\textbf{0.323}}}
				& \parbox{1.5cm}{\cellcolor{orange!40}\centering{0.911}} & \parbox{1.5cm}{\cellcolor{orange!40}\centering{0.290}}
				& \parbox{1.5cm}{\cellcolor{orange!30}\centering{0.906}} & \parbox{1.5cm}{\cellcolor{orange!30}\centering{0.274}}
				& \parbox{1.5cm}{\cellcolor{orange!20}\centering{0.904}} & \parbox{1.5cm}{\cellcolor{orange!20}\centering{0.263}}\\ 
				
				\parbox{4.2cm}{\textbf{modified residue}}
				& \parbox{1.5cm}{\cellcolor{orange!50}\centering{\textbf{0.937}}} & \parbox{1.5cm}{\cellcolor{orange!50}\centering{\textbf{0.144}}}
				& \parbox{1.5cm}{\cellcolor{orange!30}\centering{0.932}} & \parbox{1.5cm}{\cellcolor{orange!30}\centering{0.124}}
				& \parbox{1.5cm}{\cellcolor{orange!20}\centering{0.925}} & \parbox{1.5cm}{\cellcolor{orange!20}\centering{0.111}}
				& \parbox{1.5cm}{\cellcolor{orange!10}\centering{0.923}} & \parbox{1.5cm}{\cellcolor{orange!10}\centering{0.107}}
				& \parbox{1.5cm}{\cellcolor{orange!40}\centering{0.932}} & \parbox{1.5cm}{\cellcolor{orange!40}\centering{0.134}}\\ 
				
				\parbox{4.2cm}{\textbf{transmembrane domain}}
				& \parbox{1.5cm}{\cellcolor{orange!10}\centering{0.975}} & \parbox{1.5cm}{\cellcolor{orange!10}\centering{0.734}}
				& \parbox{1.5cm}{\cellcolor{orange!20}\centering{0.99}} & \parbox{1.5cm}{\cellcolor{orange!20}\centering{0.886}}
				& \parbox{1.5cm}{\cellcolor{orange!30}\centering{0.991}} & \parbox{1.5cm}{\cellcolor{orange!30}\centering{0.896}}
				& \parbox{1.5cm}{\cellcolor{orange!40}\centering{0.992}} & \parbox{1.5cm}{\cellcolor{orange!40}\centering{0.9}}
				& \parbox{1.5cm}{\cellcolor{orange!50}\centering{\textbf{0.993}}} & \parbox{1.5cm}{\cellcolor{orange!50}\centering{\textbf{0.9}}}\\ \midrule 	
				
		\end{tabular}}
	\end{table*}
	
    \subsection{Protein-Dependent Generalizability}\label{Protein-dependent generalizability}
 
\begin{figure}[ht]%
        \centering
        \includegraphics[width=1\textwidth]{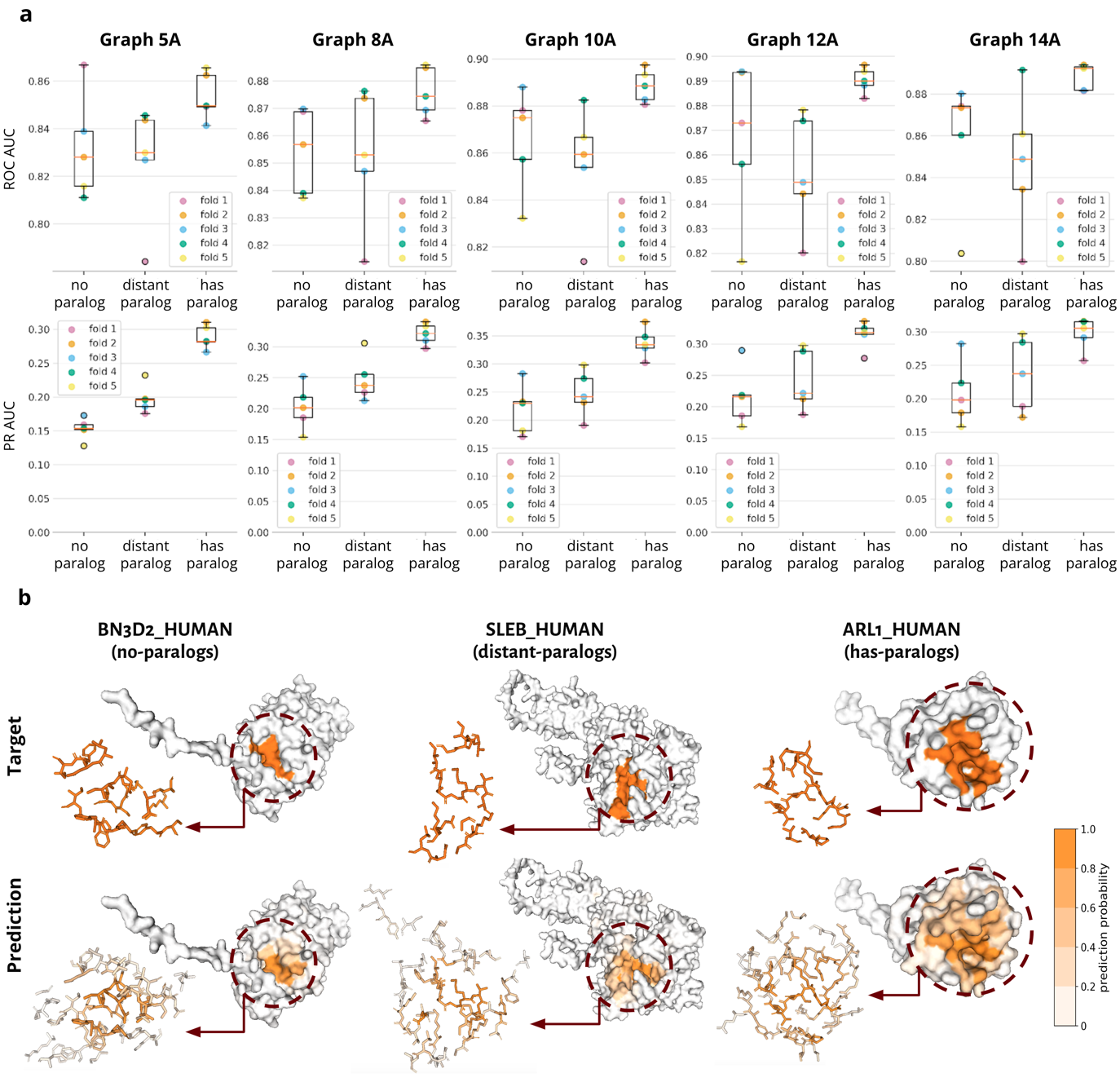}
        \caption{Protein-dependent generalizability analysis for ligand binding site prediction. ROC AUC and PR AUC distributions across validation folds obtained from three separate models at varying distance thresholds ($5\textrm{\AA}$, $8\textrm{\AA}$, $10\textrm{\AA}$, $12\textrm{\AA}$, and $14\textrm{\AA}$) and three paralog groups. (b) 3D representations of true ligand binding site (Target) and predicted ligand binding site (Prediction) of three human proteins: BN3D2-HUMAN from \textit{no-paralog} group, SLEB-HUMAN from \textit{distant-paralog} group and ARL1-HUMAN from \textit{has-paralog} group. The stick representation illustrates the main and side chain of the amino acid residues of the binding site and its surrounding. The color intensity in ``Prediction" indicates the prediction probability of the residue being classified with the downstream task. The 3D surface protein visualization and the stick representation of the amino acid residues are created using PyMOL.}\label{BLAST_Analysis}
        \end{figure}

    To evaluate NodeCoder model generalizability, test set proteins across all five cross-validation folds were binned according to their most similar training set examples.  Protein-protein similarity was defined at the sequence level and evaluated by building a BLAST database for each of the five training datasets, and performing blastp searches for each test set protein in the respective fold \cite{BLAST_ncbi}.  Test set proteins with a BLAST hit larger than $30\%$ sequence identity were binned as \textit{has-paralog}.  Those with a BLAST hit between $25\%$ and $30\%$ sequence identity were binned as \textit{distant-paralog}, while those lower than $25\%$ were binned as \textit{no-paralog}.  Distance thresholds were chosen based on the longstanding convention of a \textit{Twilight Zone} for protein sequence alignments \cite{rost1999twilight}.   Fig. \ref{BLAST_Analysis} reports NodeCoder ligand binding prediction performance on each sequence distance bin.  While test set proteins with known training set homologs are easier to predict, the performance gap is less than 0.02 ROC AUC with the group without known homologs at the optimal 10$\textrm{\AA}$ distance threshold demonstrating good overall model generalizability.   Fig. \ref{BLAST_Analysis} - (b) shows NodeCoder prediction results for ligand binding site for three human proteins: BN3D2-HUMAN from \textit{no-paralog} group, SLEB-HUMAN from \textit{has-paralog} group and ARL1-HUMAN from \textit{has-paralogs} groups. These results are obtained from models trained with $10 \textrm \AA$ models. The color intensity in ``Prediction" indicates the prediction probability of the residue being classified as a ligand binding site. For all three proteins, it can be seen that the ligand binding sites align well visually with the positively predicted residues labels, despite weaker predictive signals in the peripheral rim regions of the binding site.  Weaker prediction probabilities at the periphery contribute to the high false negative rates and consequently lower PR-AUC values.  Most datasets in this study have a large class imbalance (see supplemental table \ref{Summary_of_Folds}), leading to predictive models that are skewed to the negative class.

    \subsection{Pocket Detection Benchmarking}\label{Benchmarking Section}
    
\begin{figure}[ht]%
		\centering
		\includegraphics[width=1\textwidth]{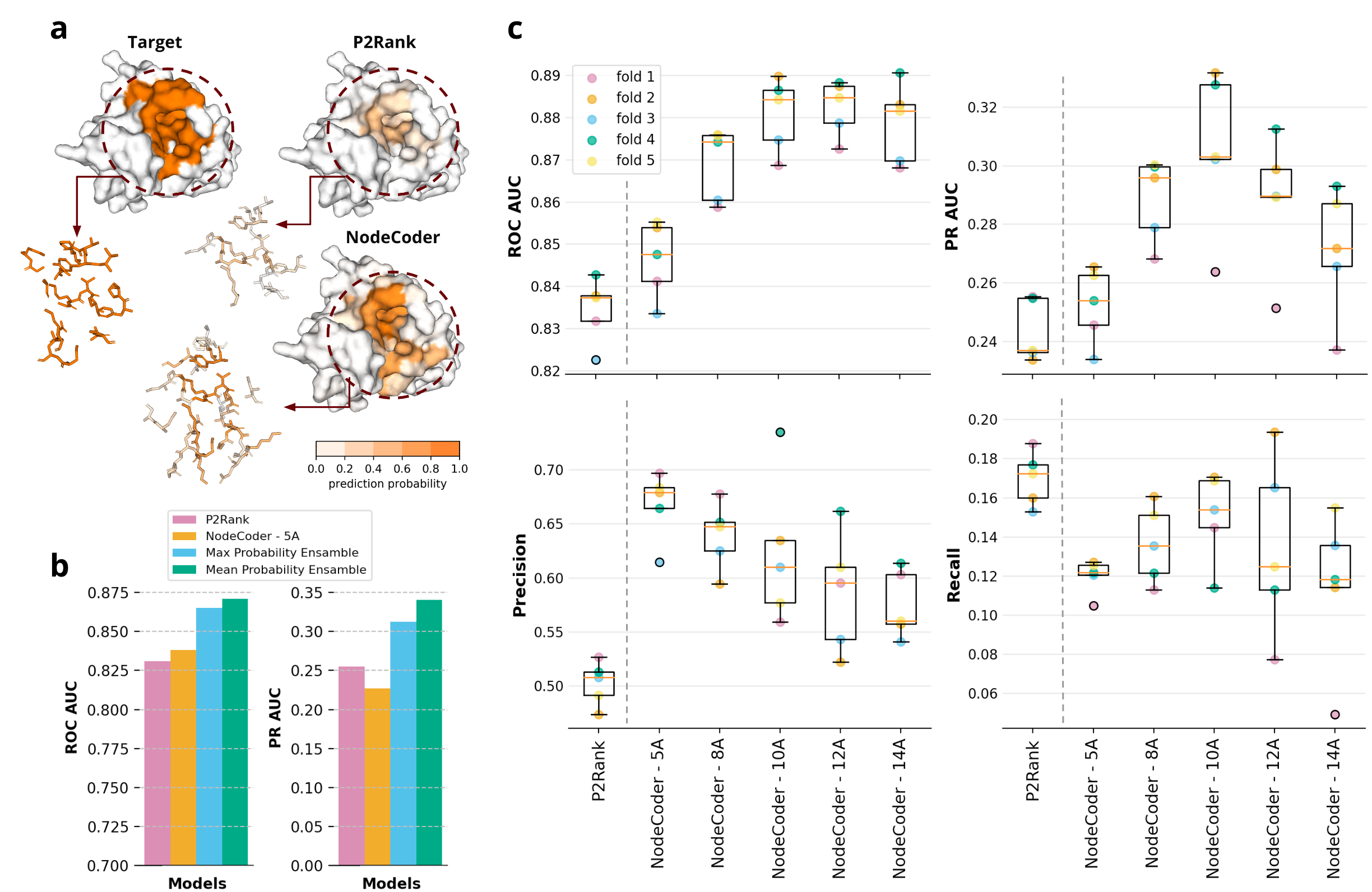}
		\caption{Comparing NodeCoder with P2RANK for ligand binding site prediction. (a) Ligand binding site of RHOB-HUMAN protein, predicted with P2Rank and the $10 \textrm{\AA}$ NodeCoder model, as well as the ``Target" set of residues known to bind small molecule ligands. For the P2Rank and NodeCoder structure images, color intensity denotes prediction confidence.  A stick representation is also provided to visualize side chains conformations. All protein visualizations are created using PyMol. (b) Performance measures of ensemble approaches with P2Rank and NodeCoder - $5 \textrm{\AA}$ models on fold-1 dataset. (c) Distribution of ROC AUC, PR AUC, $Precision$ and $Recall$ for five validation folds that are obtained from different models trained by P2RANK and NodeCoder with five distance thresholds of $5\textrm{\AA}$, $8\textrm{\AA}$ and $10\textrm{\AA}$, $12\textrm{\AA}$, $14\textrm{\AA}$.}
		\label{Benchmarking}
	\end{figure}

    To further assess NodeCoder's task-independent architecture, we compared its ligand binding residue prediction model with P2Rank, a widely-used open-source software package to detect ligand binding sites from protein structure \cite{krivak2018p2rank}.  Specifically, we used P2Rank to identify pockets and label their respective residues to perform the side-by-side comparison.  The NodeCoder model outperformed P2Rank according to the ROC AUC, PR AUC, and precision metrics, while P2Rank had a better recall performance, Fig. (\ref{Benchmarking})-(c).  An ensemble model combining both predictions was even more predictive than either algorithm alone, Fig. \ref{Benchmarking}-(b), suggesting that both approaches learn orthogonal information.  A similar interpretation can be concluded from the precision and recall behaviours observed in Fig. (\ref{Benchmarking})-(c). Specifically, P2Rank's higher recall and lower precision rate is consistent with a topology-driven approach enumerating possible pockets.  Conversely, NodeCoder's false negative tendencies are consistent with an approach that learns exclusively from experimentally-observed binding sites.  Figure (\ref{Benchmarking})-(a) illustrates the ligand binding site of RHOB-HUMAN protein, predicted with P2Rank and NodeCoder - $10 \textrm{\AA}$ model, as well as the ``Target" set of residues known to bind small molecule ligands.  While both models detect the ligand binding site, the NodeCoder model has better localization of the ligand binding site and prediction confidence in this example.  Lastly, there may be an additional opportunity to improve NodeCoder's binding site predictions by consolidating residue-level predictions with their local neighbors do define interaction patches.

    \section{Methods}\label{Methods}
    
    Fig. \ref{Schematic Representation} provides a schematic representation of the NodeCoder framework architecture for residue characterization on AlphaFold2 protein structures as the network input. First the initial representation of protein graph data is defined by assigning features to nodes and edges. Then, the GCN layers encode the information on the structure of the graph and exploits this information to update the initial representation of nodes. Next, the output prediction layer performs a learning task of node classification by employing the encoded protein graph representations obtained as output from the GCN layers. In the following sections we explain the details of graph data preparation and the NodeCoder framework.
	
    \subsection{Graph Representation of Proteins}
    \label{Protein graph}
 
    Protein residue graphs, $\mathcal{G}(\mathcal{V},\mathcal{E},A)$, are constructed from each predicted protein structure in the AlphaFold2 human proteome, represented by the adjacency matrix, $A \in \mathbb{R}^{N \times N}$ encoding connections between $N$ residues, a residue-level feature matrix, $\mathcal{V} \in \mathbb{R}^{N \times M}, \mathcal{V}=\{v_1, v_2, ..., v_N\}$, and edge feature matrix, $\mathcal{E} \in \mathbb{R}^{E \times F}$. Nodes $v_i$ correspond to individual residues and edge $e_{ij}$ correspond to inter-residue contacts within pre-set distances. The protein residue network is represented by its adjacency matrix $A$ defined as:

	\begin{align}
	    A_{ij} = 
	    \begin{cases}
	    a_{ij}        & \text{if} \; e_{ij} \in \mathcal{E} \\
	    0             & otherwise.
	    \end{cases}
	\end{align}

    $A$ is symmetric if the edges are undirected, where $A_{ij} = A_{ji} > 0$ if node $v_i$ and node $v_j$ share an edge, otherwise $A_{ij} = A_{ji} = 0$. $a_{ij} = 1$ if the graph is unweighted. We add self-connections to the adjacency matrix by converting all diagonal elements of $A$ to $1$, and form $\tilde{A} = A + I$, which indicates the output of a node in a hidden layer depends on itself and its neighbors. In the protein graph $\mathcal{G}$, an edge $e_{ij}$ connects two nodes $(v_i, v_j)$ if the residues are within a threshold distance.  Inter-residue distance is defined by the Euclidean distance between their respective $C_\alpha$ atoms. 

    \subsection{Node Features} 
    Node features encode biophysical properties of the amino acid residues within their predicted structures. Residues (nodes) are annotated with $46$ features derived directly from the 3D structure.  These include $20$ individual features for one-hot encoding amino acid type, four backbone angles (phi, psi, tau, and theta), solvent accessibility of the residue’s backbone and side chain atoms, secondary structure, backbone hydrogen bonding, positional information, side chain orientations, and structure prediction confidence (ie, pLDDT values). The full list of features is described in Table \ref{node features}. PDB file processing was performed with Biopython \cite{hamelryck2003pdb, cock2009biopython}. Solvent accessibility features were generated using FreeSASA \cite{mitternacht2016freesasa}. Secondary structure and backbone hydrogen bonding interaction features were generated with DSSP \cite{Kabsch1983, Touw2014, dssp4}.
    Features are intentionally restricted to residue properties derived exclusively from the 3D structure to ensure that the GCN model can be applied to future protein structures with a single model file input. Moreover, only structure files derived from the AlphaFold2 database are used for training to avoid source bias.
	
    \subsection{Edge Features}\label{Edge Features}
    The neighborhood of a node used in the convolution operator is the set of closest residues as determined by the distance threshold between their $C_\alpha$ atoms. The spatial relationships between residues are represented as features of the edges that connect them.  An edge weight feature denoting inter-residue distance was also introduced to capture additional spatial information in the network.  Specifically, the inverse square of the Euclidean distance between $C_\alpha$ atoms was chosen as the edge weight feature after experimenting with several distance metrics and transformations based on Euclidean, cosine, and primary sequence distances during model development.
	
    \subsection{Residue Annotations (Tasks)}
    NodeCoder models are trained to learn and predict residue-level datasets, which may be derived from 3D structures or alternative sources (e.g. sequencing or Mass spectrometry). In this study, two tasks were derived from UniProt sequence annotations: transmembrane domain residues and post-translational modifications derived from functional studies.  Four tasks were from BioLip, which indexes observed binding sites from macromolecular structures in the PDB \cite{yang2012biolip}: Small molecule ligands binding sites, nucleic acid binding sites, peptide binding sites, and metal ion binding sites.  All BioLip-documented binding sites were mapped onto their protein's UniProt canonical sequence.  The residue-level dataset contains one sequence representation per protein and each element (residue) is labeled as a binding residue if a corresponding interaction is observed in \textit{any} experimental PDB structure. Models lacking any binding annotations are not used for training, as to prevent excessive false negative labels from biasing the dataset.
	
    \subsection{Graph Convolutional Network}\label{GCN}
    NodeCoder encodes protein graph representations with a multi-layer GCN network architecture.  Each residue is encoded as a target node in the GCN, whose representation is updated by aggregating information from its neighboring residues in the protein graph. While the graph structure is shared over layers, GCN performs a series of message passing to exchange information between residues and to generate encodings that can predict the residue label. NodeCoder was designed with a multi-layer GCN architecture, first, because GCN is a suitable method for encoding proteins' features by taking into account the 3D graph-based structure of residues contact network, second, by stacking $L$ layers of GCN in network architecture, the target node representation aggregates the features of nodes whose distance is equal to $L$ from the target node, as a result of recursive neighborhood diffusion, underlined by Dwivedi et al. \cite{dwivedi2020benchmarking}.

    The NodeCoder Graph Convolutional Network is built using the PyTorch geometric library \cite{paszke2019pytorch}. A schematic representation of the model architecture is shown in Fig. \ref{Schematic Representation}.  The network encoder consists of multiple graph convolutional blocks of different sizes which are connected by ELU activation function \cite{ELU} and followed by the output layer. The output layer, consists of multiple independent GCN blocks corresponding to multiple tasks of interest, each followed by a regularizer block, a $50\%$ dropout function, to prevent overfitting. Finally, every output layer is followed by a LogSoftMax function \cite{logsoftmax} that calculates the logarithmic probability of the target of interest being a positive label. NodeCoder implements a GCN model architecture whose aggregation function is an isotropic operation, such that features of neighboring nodes are contributing equally to aggregation. NodeCoder also encodes node centrality by adding node degree to the node features' vector as the input to capture the node importance in the graph \cite{ying2021transformers}. 

    For all experiments performed, the structure of the input encoder network consists of four GCN hidden layers operating with kernels of sizes $(46\times 38)$, $(38\times 28)$, $(28\times 18)$, $(18\times 8)$, and they are followed by output layer with GCN kernels of size $(8 \times 2)$, where each output kernel correspond to predicting a single annotation.
	
    The GCN was modeled after Kipf et al. \cite{kipf2016semi}, defining the following layer-wise propagation rule for graph convolution: 
	\begin{equation}
	    H^{(l+1)} = \sigma(\tilde{D}^{-\frac{1}{2}}\tilde{A}\tilde{D}^{-\frac{1}{2}} H^{(l)} W^{(l)}),
	    \label{propagation rule}
	\end{equation}
	where, $H^{(l)} \in \mathbb{R} ^{N\times M_l}$ is the matrix of activations in layer $l$, $W^{(l)} \in \mathbb{R}^{M_{l} \times M_{l+1}}$ is the trainable weight matrix of hidden layer $l$ with $M_{l}$ feature maps. $H^{(0)} = \mathcal{V} \in \mathbb{R} ^{N\times M}$ is a matrix of input node feature vectors with $N$ nodes and $M$ input node features.
	$\sigma(\cdot)$ is a non-linear point-wise activation function like Exponential Linear Unit, $ELU(\cdot)$.
	$\tilde{A}$ is the adjacency matrix of the undirected graph $\mathcal{G}$ with added self-connections. $\hat{D}$ is the diagonal node degree matrix of $\tilde{A}$, where the diagonal elements are the degree of each node in the graph, counting the number of edges for the corresponding node $i$: $\tilde{D}_{ii} =\sum _{j} A_{ij}$. When $\tilde{A}$ is multiplied by $\tilde{D}^{-\frac{1}{2}}$, at a high level, the aggregated features from the node itself and its neighbors are averaged to normalize $\tilde{A}$ such that the scale of the output feature vectors is sustained.
	
    For residue-based classification on protein graph, with a $L$-layer GCN model, $\hat{A} = \tilde{D}^{-\frac{1}{2}}\tilde{A}\tilde{D}^{-\frac{1}{2}}$ is first calculated. Then the latent representation of each hidden layer is calculated by encoding the graph structure with the propagation rule given in equation (\ref{propagation rule}). GCN kernels were again selected for the output prediction layer.  NodeCoder models have $T$ independent output GCN blocks corresponding to $T$ annotations or tasks. The output GCN blocks take in the encoded protein graph representation obtained from the multi-layer GCN model and finally, the forward model is formed as:
    \begin{equation}
	    Z_{t} = f(\mathcal{V}, \hat{A}) = LogSoftMax(\hat{A} \; ELU(\hat{A} H^{(L)} W^{(L)}) \; W_{t}^{(L+1)}), 
	\end{equation}
	where, $t \in \{1, 2, ..., T\}$, $W_{t}^{(L+1)}$ is the weight matrix of the output GCN layer corresponding to task $t$, and $ELU(\hat{A} H^{(L)} W^{(L)})$ is the encoded protein graph obtained from the last hidden layer of the input $L$-layer GCN model. LogSoftMax is an activation function that encodes log-likelihood or logarithmic probability distribution, and its applied row-wise as:
    \begin{equation*}
	    LogSoftMax(x_{i}) = log(\frac{exp(x_{i})}{\sum_{i}exp(x_{i})}),
	\end{equation*}
	and the $ELU$ activation function is defined as:
	\begin{align*}
	    ELU(x) = 
	    \begin{cases}
	    x                   & \text{if} \; x > 0 \\
	    \alpha (e^{x}-1)    & \text{if} \; x\leq 0.
	    \end{cases}
	\end{align*}

        \subsection{Model Training}
        \label{Train}
        NodeCoder graph-based networks are trained with the torch negative log-likelihood function, NLLLoss, $\texttt{torch.nn.NLLLoss}$. The ADAM optimizer \cite{kingma2014adam} with learning rate of $0.01$, parameters $\beta_{1} = 0.9$ and $\beta_{2} = 0.999$ is chosen to minimize training loss. For the reported analysis here, all proteins are considered in one batch for model training. NodeCoder; however, can split data into multiple clusters to reduce memory usage when it is required. In this case, the clustering approach, which is performed on generated graph data, splits graph data such that an entire protein graph is in one cluster. We trained the NodeCoder's models for $10,000$ epochs and the optimum epoch is selected based on the highest PR AUC score.
        

	\subsection{Cross-Validation and Performance Metrics}
	\label{Cross-Validation}
  
	All testing and training examples in this study were obtained from human proteome, further limiting datasets to one single 3D structure model per protein. Proteins lacking any binding annotations are excluded as the excessive false negative labels could result in biasing the dataset. Consequently, there are $3823$ human proteins in our training dataset. The dataset was randomly split into five folds for cross-validation (see supplemental table \ref{Summary_of_Folds}). All performances reported in this study are averaged over the five test-train splits. Models were evaluated using receiver operating characteristic (ROC) and precision-recall (PR) curves due to the substantial imbalance of positives and negatives among datasets. In addition to ROC AUC and PR AUC scores, we also calculate $Precision$ and $Recall$ scores, respectively written as:
	\begin{equation*}
	    Precision_{t} = \frac{TP_{t}}{TP_{t}+FP_{t}},
	\end{equation*}
	\begin{equation*}
	    Recall_{t} = \frac{TP_{t}}{TP_{t}+FN_{t}},
	\end{equation*}
	for each decision threshold $t$. 
    
	\section{Discussion}\label{Discussion}

    Protein fold topology is widely recognized as the strength of AlphaFold2 predictions, whereas domain orientations and side chain conformations predictions remain a challenge \cite{Terwilliger2022.11.21.517405}.  Residue contact networks capture protein fold topology independent of side-chain conformations.  Designing a learning strategy around this protein structure abstraction focuses on the strength of high-performance protein structure prediction models, such as AlphaFold2.  In this study, we introduced a protein learning framework that augments biophysical prediction problems with structural context using predicted protein structures.  The use of a single machine learning architecture to model multiple tasks has several notable operational advantages.  For one, new models can be readily trained from any residue-mapped datasets now that reasonably reliable protein structure representations exist for nearly every sequenced protein.  This convenience facilitates experimentation with new types of tasks predictions that are typically limited in structural information or not typically structurally enabled, such as transit peptide predictions, sub-cellular localization predictions, disorder prediction, antigenicity, or more detailed post-translational modification models (phosphorylation, glycosylation, etc.).  Moreover, modeling multiple tasks with a single, unified framework allows them to simultaneously benefit in parallel from subsequent iterative improvements, such as improved features, model architectures or improvements in cross-validation strategies.  The NodeCoder software is made freely available to support open science and the development of new predictive models.

	\bmhead{Data and Code availability}
    NodeCoder software and trained models are available at \href{https://github.com/NasimAbdollahi/NodeCoder}{https://github.com/NasimAbdollahi/NodeCoder}, or readily installable via pypi at \href{https://pypi.org/project/NodeCoder/}{https://pypi.org/project/NodeCoder/}.  The open-source software includes source code to build datasets, train models and perform model predictions.  Due to the large size of the data, our training, validation, and test data splits are not available from our GitHub page, but are available from the authors upon reasonable request. 
	
	\bibliographystyle{sn-standardNature}
	\bibliography{References}
	
	\section*{Acknowledgments}
	 We thank Robert Vernon, Andreas Windemuth, Shoshana Wodak, Andrew Brereton, David Kuter, and Vijay Shahani for their insightful comments. We thank Steve Constable for his technical support while working on open-source software. This research was supported by the Mitacs Accelerate Postdoctoral Fellowship and industry partner Cyclica.
	
	\bmhead{Competing interests}
	S.S.M.K., A.M., and J.H. are employees of Cyclica, and may own stock in Cyclica Inc.
	
	\bmhead{Author contributions}
	N.A. and S.S.M.K. wrote the manuscript with input from all the authors. 
	S.S.M.K. collected protein structures and annotations from AlphaFold, UniProt, and BioLip, and worked on the prepossessing step. 
	A.M. and N.A. worked on model development. 
	N.A. developed and maintained the NodeCoder software, python package, and GitHub repository. 
	A.M. and S.S.M.K. oversaw methods and software development.
	N.A., A.M., S.S.M.K., and B.W. conceived the study and designed the experiments.
	N.A. conducted experiments and analyses.
	J.H. prepared P2Rank prediction results.
	A.M. conducted ensemble analyses. 
	B.W., S.S.M.K., and A.M. contributed to the analysis and discussion of the results.
	N.A. worked on figure design. 
	N.A. and S.S.M.K. worked on PyMOL visualizations. 
	S.S.M.K. and B.W. supervised the research. 
	All authors reviewed the manuscript and approved it for submission.
	
	\newcommand{\beginsupplement}{%
        \setcounter{table}{0}
        \renewcommand{\thetable}{S\arabic{table}}%
        \setcounter{figure}{0}
        \renewcommand{\thefigure}{S\arabic{figure}}%
     }
    
    \beginsupplement
    \section*{Supplementary Materials}\label{Supplementary}
    \subsection*{Extended Data}\label{Extended data}
    Table \ref{Summary_of_Folds} 
    reports the total node count and percentage of positive labels in each train and validation fold. Table \ref{node features} describes node features. Table \ref{annotation list} includes the annotation tasks used in evaluating NodeCoder and reported analysis. 

	\begin{table*}[!h]
	\centering
	\caption{Summary of total node count and percentage of positive labels in each train and validation fold.}
	\label{Summary_of_Folds}
	\scalebox{0.5}{\begin{tabular} {@{}ccccccccccccl@{}}
			\midrule 
			\parbox{4.2cm}{\textbf{cross-validation fold:}} & \multicolumn{2}{c}{\parbox{2cm}{\centering \textbf{fold 1}}} & \multicolumn{2}{c}{\parbox{2cm}{\centering \textbf{fold 2}}} & \multicolumn{2}{c}{\parbox{2cm}{\centering \textbf{fold 3}}} & \multicolumn{2}{c}{\parbox{2cm}{\centering \textbf{fold 4}}} & \multicolumn{2}{c}{\parbox{2cm}{\centering \textbf{fold 5}}}\\ \midrule
			
			\parbox{1cm}{\centering \textbf{}} & \parbox{1.5cm}{\centering \textbf{train}} & \parbox{1.5cm}{\textbf{validation}} & \parbox{1.5cm}{\centering \textbf{train}} & \parbox{1.5cm}{\textbf{validation}} & \parbox{1.5cm}{\centering \textbf{train}} & \parbox{1.5cm}{\textbf{validation}} & \parbox{1.5cm}{\centering \textbf{train}} & \parbox{1.5cm}{\textbf{validation}} & \parbox{1.5cm}{\centering \textbf{train}} & \parbox{1.5cm}{\textbf{validation}} \\ \midrule
			
			\parbox{4.2cm}{\textbf{protein count:}} & \parbox{1.5cm}{\centering \text{3058}} & \parbox{1.5cm}{\centering \text{765}} & \parbox{1.5cm}{\centering \text{3058}} & \parbox{1.5cm}{\centering \text{765}} & \parbox{1.5cm}{\centering \text{3058}} & \parbox{1.5cm}{\centering \text{765}} & \parbox{1.5cm}{\centering \text{3058}} & \parbox{1.5cm}{\centering \text{765}} & \parbox{1.5cm}{\centering \text{3060}} & \parbox{1.5cm}{\centering \text{763}} \\ \midrule
			
			\parbox{4.2cm}{\textbf{node count:}} & \parbox{1.5cm}{\centering \text{1,732012}} & \parbox{1.5cm}{\centering \text{440958}} & \parbox{1.5cm}{\centering \text{1,727380}} & \parbox{1.5cm}{\centering \text{445590}} & \parbox{1.5cm}{\centering \text{1,740995}} & \parbox{1.5cm}{\centering \text{431975}} & \parbox{1.5cm}{\centering \text{1,756966}} & \parbox{1.5cm}{\centering \text{416004}} & \parbox{1.5cm}{\centering \text{1,734527}} & \parbox{1.5cm}{\centering \text{438443}} \\ \midrule
			
			\parbox{4.2cm}{\textbf{ligand binding site}} 
			& \parbox{1.5cm}{\centering{2.436}} & \parbox{1.5cm}{\centering{2.453}}
			& \parbox{1.5cm}{\centering{2.455}} & \parbox{1.5cm}{\centering{2.376}}
			& \parbox{1.5cm}{\centering{2.407}} & \parbox{1.5cm}{\centering{2.568}}
			& \parbox{1.5cm}{\centering{2.422}} & \parbox{1.5cm}{\centering{2.509}}
			& \parbox{1.5cm}{\centering{2.475}} & \parbox{1.5cm}{\centering{2.295}}\\ \midrule 	
			
			\parbox{4.2cm}{\textbf{peptide binding site}}
			& \parbox{1.5cm}{\centering{0.976}} & \parbox{1.5cm}{\centering{0.983}}
			& \parbox{1.5cm}{\centering{0.995}} & \parbox{1.5cm}{\centering{0.910}} 
			& \parbox{1.5cm}{\centering{0.991}} & \parbox{1.5cm}{\centering{0.925}} 
			& \parbox{1.5cm}{\centering{0.966}} & \parbox{1.5cm}{\centering{1.029}} 
			& \parbox{1.5cm}{\centering{0.961}} & \parbox{1.5cm}{\centering{1.044}}\\ \midrule 	
			
			\parbox{4.2cm}{\textbf{nucleic acid binding site}}
			& \parbox{1.5cm}{\centering{0.623}} & \parbox{1.5cm}{\centering{0.505}} 
			& \parbox{1.5cm}{\centering{0.604}} & \parbox{1.5cm}{\centering{0.578}} 
			& \parbox{1.5cm}{\centering{0.593}} & \parbox{1.5cm}{\centering{0.625}} 
			& \parbox{1.5cm}{\centering{0.577}} & \parbox{1.5cm}{\centering{0.690}} 
			& \parbox{1.5cm}{\centering{0.598}} & \parbox{1.5cm}{\centering{0.604}}\\ \midrule 	
			
			\parbox{4.2cm}{\textbf{metal ion binding site}}
			& \parbox{1.5cm}{\centering{0.755}} & \parbox{1.5cm}{\centering{0.680}} 
			& \parbox{1.5cm}{\centering{0.745}} & \parbox{1.5cm}{\centering{0.719}} 
			& \parbox{1.5cm}{\centering{0.712}} & \parbox{1.5cm}{\centering{0.852}}
			& \parbox{1.5cm}{\centering{0.737}} & \parbox{1.5cm}{\centering{0.753}} 
			& \parbox{1.5cm}{\centering{0.751}} & \parbox{1.5cm}{\centering{0.697}}\\ \midrule 	
			
			\parbox{4.2cm}{\textbf{modified residue}}
			& \parbox{1.5cm}{\centering{2.008}} & \parbox{1.5cm}{\centering{2.095}} 
			& \parbox{1.5cm}{\centering{2.046}} & \parbox{1.5cm}{\centering{1.946}} 
			& \parbox{1.5cm}{\centering{2.056}} & \parbox{1.5cm}{\centering{1.903}}
			& \parbox{1.5cm}{\centering{2.012}} & \parbox{1.5cm}{\centering{2.083}} 
			& \parbox{1.5cm}{\centering{2.005}} & \parbox{1.5cm}{\centering{2.102}} \\ \midrule 	
			\parbox{4.2cm}{\textbf{transmembrane domain}}
			& \parbox{1.5cm}{\centering{0.766}} & \parbox{1.5cm}{\centering{0.716}} 
			& \parbox{1.5cm}{\centering{0.755}} & \parbox{1.5cm}{\centering{0.759}}
			& \parbox{1.5cm}{\centering{0.738}} & \parbox{1.5cm}{\centering{0.829}}
			& \parbox{1.5cm}{\centering{0.764}} & \parbox{1.5cm}{\centering{0.722}} 
			& \parbox{1.5cm}{\centering{0.757}} & \parbox{1.5cm}{\centering{0.752}}\\ \midrule 	
				
		\end{tabular}}
	\end{table*}

		\begin{table*}[!h]
		\centering
		\caption{Node features.}
		\label{node features}
		\scalebox{0.7}{\begin{tabular} {@{}cccl@{}}
				\midrule 
				\parbox{4cm}{\textbf{feature name}} & \parbox{7cm}{\textbf{description}} & \parbox{2cm}{\centering{\textbf{count}}} \\ \midrule
				
				\parbox{4cm}{\textbf{Amino Acid}} & \parbox{7cm}{Twenty binary features to encode the node's corresponding amino acid (one-hot encoding)} & \parbox{1cm}{\centering{20}} \\ \midrule 	
				
				\parbox{4cm}{\textbf{iPlus}} & \parbox{7cm}{Amino acid immediately prior to this node in the protein's primary sequence} & \parbox{1cm}{\centering{1}} \\ \midrule 	
				
				\parbox{4cm}{\textbf{iMinus}} & \parbox{7cm}{Amino acid immediately following this node in the protein's primary sequence} & \parbox{1cm}{\centering{1}} \\ \midrule 	
				
				\parbox{4cm}{\textbf{DSSP}} & \parbox{7cm}{Seven features encoding DSSP secondary structure assignment (one-hot encoding) and eight features that encode DSSP H-bonding interactions} & \parbox{1cm}{\centering{15}} \\ \midrule 	
				
				\parbox{4cm}{\textbf{dihedral angles}} & \parbox{7cm}{dihedral angles: $\phi , \psi , \tau , \theta $} & \parbox{1cm}{\centering{4}} \\ \midrule 	
				
				\parbox{4cm}{\textbf{BBSASA}} & \parbox{7cm}{Back Bone Solvent Accessibility} & \parbox{1cm}{\centering{1}} \\ \midrule 	
				
				\parbox{4cm}{\textbf{SCSASA}} & \parbox{7cm}{Side Chain Solvent Accessibility} & \parbox{1cm}{\centering{1}} \\ \midrule 	
				
				\parbox{4cm}{\textbf{pLDDT}} & \parbox{7cm}{AlphaFold per-residue confidence metric \cite{AlphaFold2}} & \parbox{1cm}{\centering{1}} \\ \midrule 	
				
				\parbox{4cm}{\textbf{centric distance}} & \parbox{7cm}{euclidean distance of amino acid residue from the center of protein} & \parbox{1cm}{\centering{1}} \\ \midrule 	
				
				\parbox{4cm}{\textbf{centric cosine distance}} & \parbox{7cm}{cosine distance of amino acid residue from the center of protein} & \parbox{1cm}{\centering{1}} \\ \midrule 

                    \parbox{4cm}{\textbf{centrality$^*$}} & \parbox{7cm}{node degree capturing the node importance in the graph} & \parbox{1cm}{\centering{1}} \\ \midrule 
			\multicolumn{3}{l}{\small *This feature is obtained from graph representation of protein.} \\ 
		\end{tabular}}
	\end{table*}

\begin{table*}[!h]
		\centering
		\caption{Annotation tasks used in reported analysis.}
		\label{annotation list}
		\scalebox{0.7}{\begin{tabular} {@{}cccl@{}}
				\midrule 
				\parbox{3cm}{\textbf{task name\\(annotation)}} & \parbox{7cm}{\textbf{annotation description}} & \parbox{2cm}{\textbf{mapped from}} \\ \midrule
    
				\parbox{3cm}{\texttt{y\_Ligand}} & \parbox{7cm}{ligand binding site} & \parbox{2cm}{BioLip} \\ \midrule 	
				
				\parbox{3cm}{\texttt{y\_Peptide}} & \parbox{7cm}{peptide binding site}& \parbox{2cm}{BioLip}\\ \midrule 	
				
				\parbox{3cm}{\texttt{y\_Nucleic}} & \parbox{7cm}{nucleic binding site}& \parbox{2cm}{BioLip}\\ \midrule 	
				
				\parbox{3cm}{\texttt{y\_Inorganic}} & \parbox{7cm}{metal ion binding site}& \parbox{2cm}{BioLip}\\ \midrule 	
				
				\parbox{3cm}{\texttt{y\_TRANSMEM}} & \parbox{7cm}{transmembrane domain} & \parbox{2cm}{UniProt}\\ \midrule 	
				
				\parbox{3cm}{\texttt{y\_MOD\_RES}} & \parbox{7cm}{modified residue} & \parbox{2cm}{UniProt}\\ \midrule 	
			
		\end{tabular}}
	\end{table*}

    \subsection*{Additional Results}\label{Additional Results}
\begin{figure*}[!h]%
		\centering
	\includegraphics[width=1\textwidth]{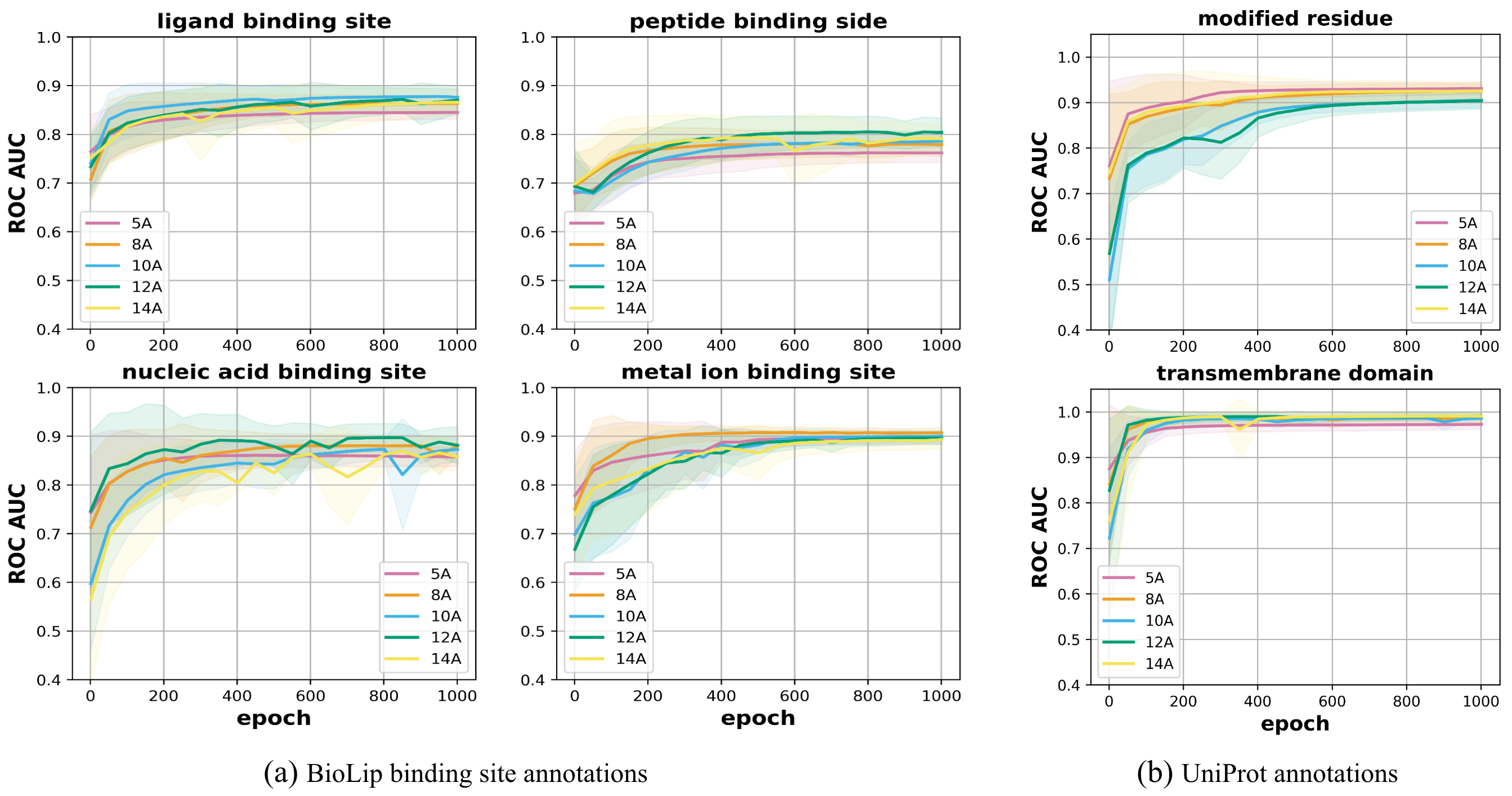}
		\caption{ROC AUC for six residue classification tasks, each generated by separate models at varying distance thresholds for the protein contact network ($5\textrm{\AA}$, $8\textrm{\AA}$, $10\textrm{\AA}$, $12\textrm{\AA}$, and $14\textrm{\AA}$). The solid line is the average on five cross-validation folds and the shaded area illustrates the average $\pm$ standard deviation. (a) Four classification tasks were obtained by mapping residue annotations from multiple BioLip entries onto the corresponding protein dataset, providing trainable labels, including ligand, peptide, ion, and nucleic acid binding sites. (b) Two classification tasks were derived from UniProt annotations, including transmembrane and modified residue.}\label{Results_PerformanceCurve_ROCAUC}
    \end{figure*}

\begin{figure*}[!h]%
		\centering
		\includegraphics[width=1\textwidth]{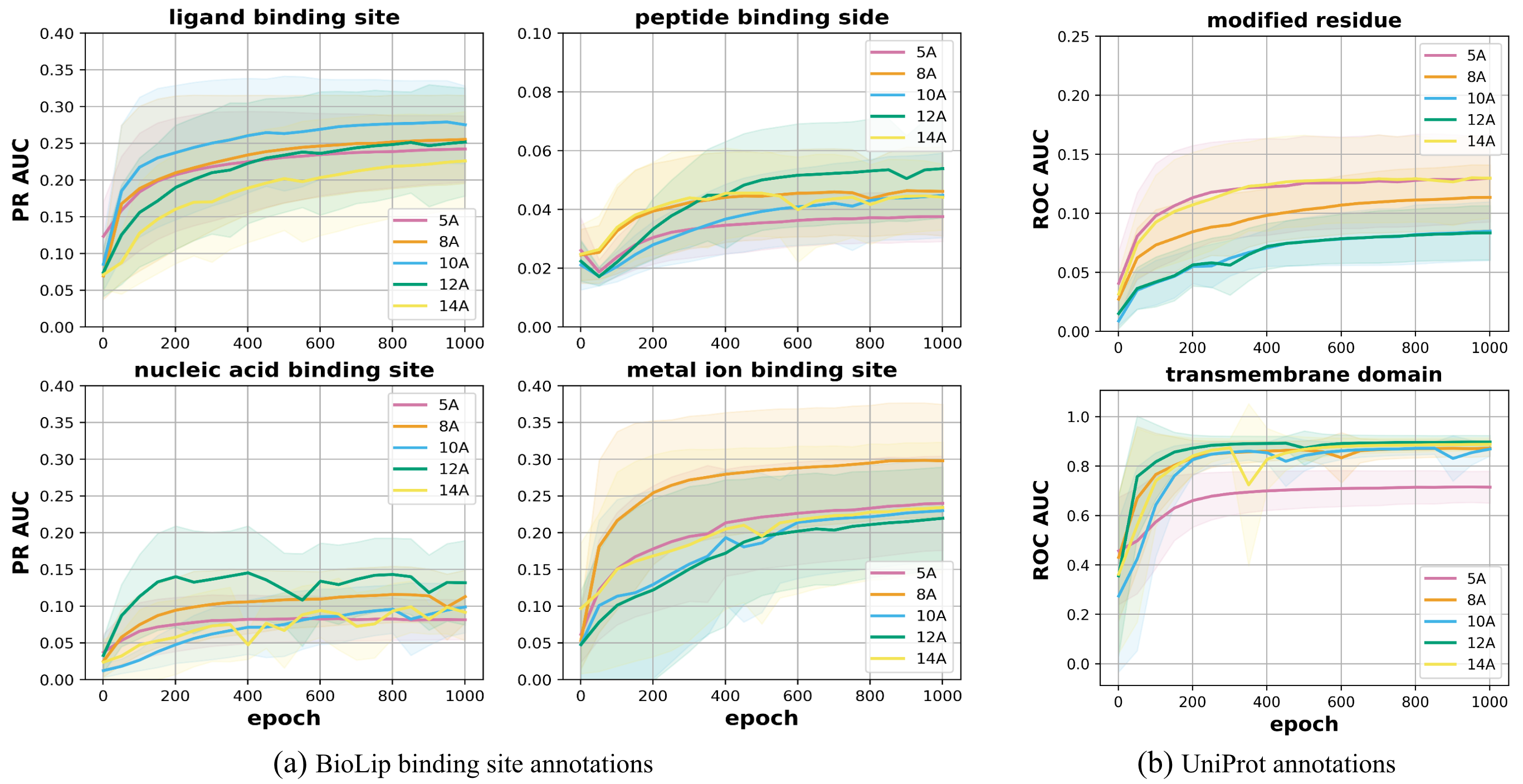}
		\caption{PR AUC for six residue classification tasks, each generated by separate models at varying distance thresholds for the protein contact network ($5\textrm{\AA}$, $8\textrm{\AA}$, $10\textrm{\AA}$, $12\textrm{\AA}$, and $14\textrm{\AA}$). The solid line is the average on five cross-validation folds and the shaded area illustrates the average $\pm$ standard deviation. (a) Four classification tasks were obtained by mapping residue annotations from multiple BioLip entries onto the corresponding protein dataset, providing trainable labels, including ligand, peptide, ion, and nucleic acid binding sites. (b) Two classification tasks were derived from UniProt annotations, including transmembrane and modified residue.}\label{Results_PerformanceCurve_PRAUC}
    \end{figure*}

    Fig. \ref{Results_PerformanceCurve_ROCAUC} and \ref{Results_PerformanceCurve_PRAUC} 
    report  performance curves obtained from training models for six predictive tasks including: ligand, peptide, ion, and nucleic acid binding sites, obtained by mapping residue annotations from multiple BioLip entries onto the corresponding protein dataset, and transmembrane domain and modified residue derived from UniProt annotations.

    \subsection*{NodeCoder Usage}\label{NodeCoder Usage}
    The NodeCoder source code and data models are available through GitHub.  Alternatively, the python package is can be installed via pypi to train new models or perform inference.  Pre-trained models can be used to predict different interaction surfaces on AlphaFold2 protein structures. The following code snippets provide instructions on how to install and run NodeCoder for train and inference (the complete guideline on how to use NodeCoder can be found at \href{https://pypi.org/project/NodeCoder/}{https://pypi.org/project/NodeCoder/}): 
    
    \subsubsection*{Installing NodeCoder}
    \begin{lstlisting}[caption=Code snippet for installing NodeCoder python package]
    $ conda create -n <your_python_env> python=3.8.5
    $ conda activate <your_python_env>
    $ pip install --extra-index-url https://pypi.org/project/ NodeCoder
    \end{lstlisting}
    
    

    
    


    

    
    \subsubsection*{Preprocessing Raw Data}
    \begin{lstlisting}[language=Python, caption=Python code snippet for preprocessing raw data]
    from NodeCoder import preprocess_raw_data

    preprocess_raw_data.main(alphafold_data_path='.', uniprot_data_path='.'
                         ,biolip_data_path='.', biolip_data_skip_path='.', 
                         TAX_ID='9606', PROTEOME_ID='UP000005640')
    \end{lstlisting}
    
    \subsubsection*{Generate Graph Data}
    \begin{lstlisting}[language=Python, caption=Python code snippet for generating protein graph data]
    from NodeCoder import generate_graph_data

    generate_graph_data.main(TAX_ID='9606', PROTEOME_ID='UP000005640',                           threshold_dist=5, cross_validation_fold_number=5)
    \end{lstlisting}
    
	\subsubsection*{Train}    
	\begin{lstlisting}[language=Python, caption=Python code snippet for training NodeCoder on generated protein graph data]
    from NodeCoder import train

    train.main(threshold_dist=5, multi_task_learning=False,
                Task=['y_Ligand'], centrality_feature=True,
                cross_validation_fold_number=5, epochs=1000)
    \end{lstlisting}
    
	\subsubsection*{Inference} 
	\begin{lstlisting}[language=Python, caption=Python code snippet for running inference with trained NodeCoder model]
    from NodeCoder import predict
    
    predict.main(protein_ID='KI3L1_HUMAN', threshold_dist=5,
                trained_model_fold_number=1, multi_task_learning=False,
                Task=['y_Ligand'], centrality_feature=True,
                cross_validation_fold_number=5, epochs=1000)
    \end{lstlisting}
	
	
		

\end{document}